\documentstyle[twocolumn,aps,prb,epsf]{revtex}

\begin{document}

\newcommand{\beq}{\begin {equation}}
\newcommand{\eeq}{\end {equation}}
\newcommand{\etal}{{\it et al.}} 
\newcommand{\bje}{z_{B}}
\newcommand{\hut}{\hat {\bf u}}
\hsize 6in
\vsize 9in 
\oddsidemargin 0in
\voffset 1.0cm

\title{A variational approach to necklaces formation in polyelectrolytes}
\author{{\bf F. J. Solis and M. Olvera de la Cruz  } \\
 	{\it Deparment of Materials Science and Engineering,} \\
	{\it Northwestern University, Evanston, Illinois 60208-3108. }}

\maketitle

\begin{abstract} By means of a variational approach we study the conditions
under which a polyelectrolyte in a bad solvent will undergo a transition
from a rod-like structure to a ``necklace'' structure in which 
the chain collapses into a series of globules joined by stretched 
chain segments.
\end{abstract}

\section{Introduction}

	It is well established that in a salt-free environment and under
good or $\Theta$-solvent conditions, the 
electrostatic interaction between the monomers of a single polyelectrolyte
chain induces a rod-like conformation. In the pressence of 
of a small amount  of salt, if the ions are homogeneously
distributed such that their interaction with the 
chain is limited to screen the elctrostatic interactions, a rod-like 
conformation is also expected.
Even in this simple model, the introduction of a a secondary interaction can
lead to interesting structural modifications. 
Dobrynin \etal \cite{Rub} have presented a scaling argument and 
computing evidence that by turning on an attractive interaction between the
monomers, the standard rod-like 
conformation \cite{deG}, is abandon in favor of a ``necklace''
one. The necklace structure consists of
highly stretched segments alternating with collapsed globules, the ``beads'' 
of the necklace. Another mean field calculation by Balazs \etal \cite{Bal} 
also produce 
similar results. In this paper we further explore this possibility,
by means of a variational approach. The techniques we use here  are adapted 
to the case of long (effectively infinite) weakly charged polymers.

The necklace structure was first suggested 
by Kantor and Kardar \cite{Kantor} in their study of  polyampholytes.
This was argued in the basis of a similarity  with the Rayleigh instability 
\cite{Rayleigh} that appears in charged  fluids. Namely, a large charged 
droplet of fluid will split into smaller droplets to put the charges far 
away from each other and the splitting is only stopped by the increasing 
cost of surface energy. The droplets are kept together by cohesional
molecular forces, which in the polymer case are replaced by the bonds between 
monomers.

\section{Variational approach to polyelectrolyte conformation}

	The variational approach has been often used in the investigation
of the structure of polyelectrolyte chains. This technique was first
introduced in this context by de Gennes \etal \cite{deG}. (Relevant work in 
this area by means of this and other techniques has been reviewed of Barrat and Joany \cite{J&B}).

	The  basic variational principle \cite{Fey} states that given the 
Hamiltonian of the system $H$, and a family of simpler trial Hamiltonians 
$H_{t}$ characterized by a set of parameters ${a,b,\ldots}$, the free energy 
of the system satisfies the inequality
\beq
	F \leq F_{t} +\langle H-H_{t}\rangle_{t} \label{basic}
\eeq
where $F_{t}$ is the free energy of the trial Hamiltonian, and the braces 
indicate averages with respect to the statistical ensemble associated with 
the trial Hamiltonian. From among the family of trial Hamiltonians 
one can then choose the one that gives the lowest upper bound for the true 
free energy. Approximations to the expectation values of 
observables of the system can be then taken to be those calculated  
by using the optimal trial Hamiltonian. 

	The usual problem with this method is that while the bound
on the free energy is rigorously true, it might be useless and misleading
if the family of Hamiltonians is not close enough (as operators in some
space) to the true Hamiltonian. Solving this problem requires the use 
of general enough trial Hamiltonians with which practical calculations 
can still be done and that from a heuristic point of view one can expect 
them to capture the essential features of the system. Further, a successful 
application of this method can be very instructive if the parameters that 
describe the trial Hamiltonians have a direct physical interpretation.
 
	The model Hamiltonian of the polyelectrolyte 
chain that we consider is the well established  (continuum version) form,
\beq
	H=\int{dn} \, \frac{3kT}{2b^{2}} \,
	\frac{d{\bf r}}{dn}{\bf\cdot}\frac{d{\bf r}}{dn} +
	\int dn \int dn' \,  V({\bf r}(n)-{\bf r}(n'))
\eeq
with the usual entropic term and with the interaction potential between 
segments $V$ modeled as having two parts: a hard-core interaction
\beq
	V_{hc}= \nu_{0} \delta({\bf r})
\eeq
that we will consider attractive (i.e. $\nu_{0} < 0$), wich corresponds
to the immersion of the ploymer in a bad solvent, 
and a screened electrostatic part
\beq
	V_{e} = f^{2}e_{0}^{2}\frac{e^{-\kappa r}}{\epsilon r},
\eeq
where $f$ is the valence per monomer, $e_{0}$ the electronic charge, 
and $\epsilon$ the permittivity of the medium. We will consider only the case in
which the charges are quenched in the polymer backbone.

	To observe the necklace conformation, the trial Hamiltonians must 
explicitly include the possibility of uniform and 
non-uniform stretching. 
We choose the following version of the stretched chain:
\beq
	H=\int dn \, \frac{3}{2b'^{2}} \,
	\left(\frac{d{\bf r}}{dn}-\frac{d{\bf r}_{0}}{dn}\right) 
	{\bf\cdot}\left(\frac{d{\bf r}}{dn}-\frac{d{\bf r}_{0}}{dn}\right) 
\eeq
This trial Hamiltonian incorporates Gaussian fluctuations 
around a background stretched conformation ${\bf r}_{0}$ with a 
renormalized Khun length $b'$. In the basic
stretched chain model, the conformation has uniform stretching but we
will consider strechings with further spatial dependence. 

	More precisely, we will consider background conformations of
the form 
\beq
	{\bf r}_{0}(n)= \rho_{0}n\hut +\epsilon\cos(qn)\hut
\eeq
corresponding to a uniform unidirectional stretching modified by small
modulations of the stretching along the same direction. This modulation in 
the average stretching ${\bf r}_{0}$
creates zones of accumulation and depletion of monomers along the 
basic rod-like conformation. We will show that for a range of values
of the basic parameters of the problem, the minimum free energy 
with a uniform stretching is larger than if we include modulations
at certain wavelengths; i.e., we will not find the optimal stretching
but prove that the rod conformation is unstable against infinitesimal 
modulation. The wavelength of the most unstable mode thus predicts the
spacing of the beads. (See figure 1).

\begin{figure}
\epsfxsize=2.5in
\epsfbox{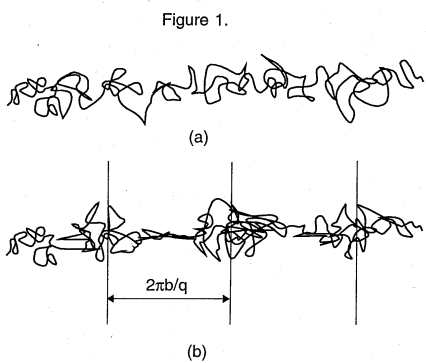}
\caption{ Typical configurations of the polyelectrolyte 
chain: (a) uniform stretching and (b) under modulated stretching
with wavelenght $2\pi b/q$.
}
\end{figure}

	Bratko and Dawson \cite{Bratko} have considered the same model 
Hamiltonian and a similar variational principle that uses 
a general Gaussian matrix of correlation between monomers and  
a constrain on the allowed 
conformations that keeps the bond-length fixed. This last constraint has 
also been incorporated in a variational approach 
by Ha and Thirumalai \cite{Ha} in their disccusion of the 
rigidity of a polyelectrolye and, as discussed
below, regularizes the behavior of the model in the strongly stretched and 
collapsed limits. In the previous study \cite{Bratko} of the  
polyelectrolyte in bad solvent a persistent stretching was not considered
({\it i.e.} $d{\bf r}_{0}/dn$ is set to $0$) and the stretching of
the chian was only observed through the large amplitudes of the 
fluctuations $\langle {\bf r}(n) {\bf r}(n') \rangle$. Instead, our 
approach takes the average stretched conformation ${\bf r}_0$ as the basic
quantity describing the system.

	The possibilities of uniform and non-uniform stretching have been 
properly  considered by Jonsson \etal \cite{Bo} when looking for the 
corrections to the stretching arising from the finite size of the 
polyelectrolyte. Their work, however, has not included consideration
of further interactions beyond the electrostatic one.

\section{Basic Rod-Like conformations}

	We start by looking at the minimum of the variational energy 
that can be obtained
by considering only trial Hamiltonians with uniform background stretchings. 
Thus, we write 
\beq
\rho(n)=\rho_0
\eeq 
and the expressions 
in (\ref{basic}) are calculated as follows. 
First, the free energy for each of the springs associated to each 
pair of neighboring monomers is, up to a constant, $-3k_{B}T\ln b'$. 
Next, the average of the stretching energy per monomer(i.e. per unit length of 
the continuous chain) of the original chain 
in this ensemble has the forced background contribution $(3/2)k_{B}\rho^{2}$
and the basic random walk part $(3/2)(b'/b)^2k_{B}T$. 

	The probability distribution for the position of two chain 
segments separated by $n$ monomer units $  G({\bf r}(s+n),{\bf r}(s))=
G({\bf r}(n)-{\bf r}(0))$, is just a Gaussian function with 
mean 
\beq \langle {\bf r}(n)-{\bf r}(0) \rangle =\rho_{0} n \hut \eeq
 and variance
\beq \langle({\bf r}(n)-{\bf f}_{0}-\rho n \hat{\bf u})^{2}\rangle =b'^{2}.
\eeq
Therefore, for an interaction  potential $V(r)$, the averaged interaction 
energy per unit length is
\beq
	\langle  V \rangle= 2\int_{0}^{\infty} ds \int d{\bf r}
	G({\bf r}(s)-{\bf r}(0))
	V(|{\bf r}(s)-{\bf r}(0)|)
\eeq
or using the potential's Fourier transform and the explicit form of the correlation
between monomers:
\begin{eqnarray}
	\langle V \rangle &=& 2\int_{0}^{\infty}ds \nonumber \\
	&&\int\frac{d{\bf k}}{(2\pi)^{3/2}}
	\exp[-\frac{k^{2}b'^{2}s}{6}+i{\bf k}\cdot\hut\rho_{0} s]
	\hat V({\bf k}).
\end{eqnarray}
Integrals of this type that diverge on the $s \rightarrow 0$ limit, are
regularized by the short distance cut-off $s_{min}=1$ (namely, one
bond distance). 

	The final result of the evaluation of the averages is 
(in units of $k_{B}T$), 
\begin{eqnarray}
	F &=& -3\ln b' +\frac{3}{2}(b'^{2}+\rho_{0}^{2}-1)
	+ \frac{6\bje}{\rho_0}\ln\left(1+\frac{6\rho_0}{b'^{2}\kappa}\right) 
	\nonumber \\
	&&+\frac{\nu}{b'^3}\left(\left(\frac{3}{2\pi}\right)^{3/2}-
	\frac{9}{4\pi}\frac{\rho}{b'}\right). \label{eval}
\end{eqnarray}
In this expression and in the rest of the paper, all distances are measured
in Kuhn length units. We have also introduced the Bjerrum
number  $\bje= f^{2}e^{2}/k_{B}T\epsilon b$, and a dimensionless 
interaction pre-factor $\nu=\nu_{0}b^{3}/k_{B}T$.

	Minimization of expression (\ref{eval}) with respect to $\rho_{0}$, 
and $b'$, gives the basic stretched conformation as a function of the 
basic parameters, $\kappa$, $\bje$ and $\nu$. In general, there exists two
minimums for this expression: one in the stretched region $\rho_{0} \neq 0$
and another in which the chain is collapsed (or Gaussian) 
$\rho_{0} \approx 0$. 
The present method and approximations, however, is suited to study 
only stretched conformations and  cannot be reliably used to decide
between the two states. For example, in our proposed variational 
family of Hamiltonians the renormalized Kuhn length $b'$ is 
intended to take values in the neighborhood of $1$ (i.e. $b' \approx b$). 
Thus, to study the globule-like collapsed chain it is  necessary to modify
the present variational principle and introduce a slightly different model Hamiltonian 
(by, for example, introducing a bond length constraint) that prevents the 
unphysical  collapse  $b'\rightarrow 0$.  To a large  extent results 
concerning this collapsed 
regime can be found in the work of Bratko and Dowson~\cite{Bratko}. 
The possibility that a collapsed state might have a  lower energy than
the stretched state leads to modifications of our results that will
be pointed out in the rest of the paper.

	In finding the minimum of the evaluation of the averages, eq. (\ref{eval}), 
we observe that for large values of the attractive interaction there is no local minimum
with finite stretching.  In figure 2 we show a numerical example of the behavior of the 
stretching and the renormalized bond length, as functions of $\nu$ and $\bje$. 
 According to  the previous 
discussion, the real stretched region in the $\nu$-$\bje$ diagram is smaller
than the one shown since the collapsed minimum will have  a lower
energy when the stretched minimum becomes unstable. 
In the examples provided we use
small values of the Bjerrum number, to keep  the stretching 
consistent with our assumptions of Gaussian springs. Stretchings
of order one should  be 
treated with a suitable representation of the 
constraint of fixed bond length. Further, in the units we use, we 
look at regions of parameter space that corresopond to fairly large 
hard core interactions 
but that still represent experimentally realizable conditions.

\begin{figure}
\epsfxsize=2.5in
\epsfbox{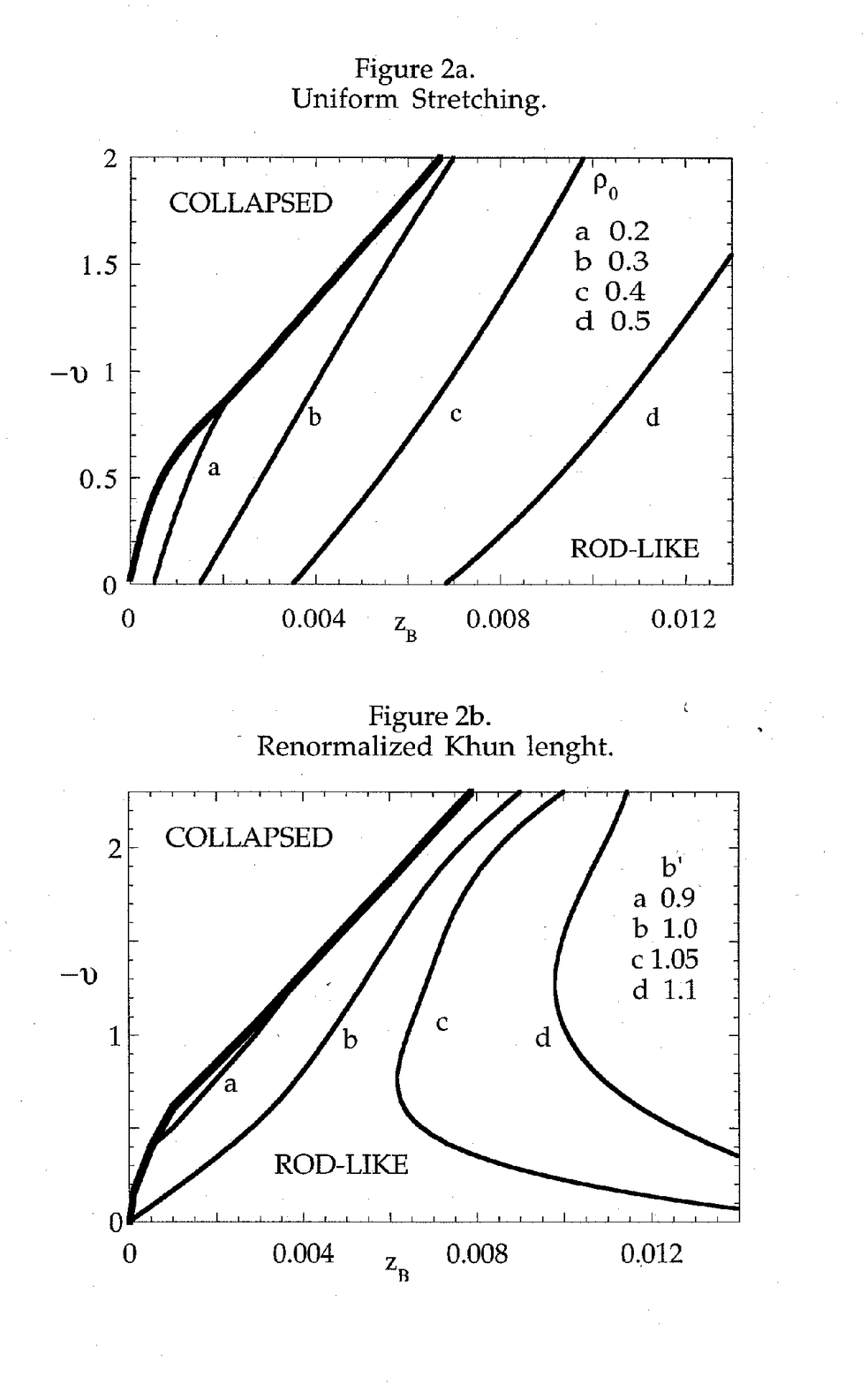}
\caption{The $-\nu-\bje$ diagram for $\kappa=10^{-4}$, where the thick 
lines mark the boundary between the
collapsed and rod-like states. Within the 
rod-like region it is shown (a) lines with constant values of the 
uniform stretching $\rho_{0}$ and (b) lines with constant values of 
the renormalized Kuhn lenght. 
}
\end{figure}

	Finally, we point out that the criteria for the validty of these 
results in the case of finite size chains is as follows.  The 
induced end to end stretching $\rho_0 N $ must be larger than the 
natural chain size $N^{1/2}$. Thus, our conclusion will be relevant
for systems satisfying the condition
\beq
	\rho_{0} \geq N^{-1/2}
\eeq

\section{Stability of rod-like conformations}

	We will now determine the stability
of the solutions found for uniform stretching. For this, we introduce a 
small perturbation on the background stretching, with a well defined 
wavenumber $q$ measured along the backbone:
\beq
	\rho(s)= \rho_{0} s + \rho_{q}\cos(qs). \label{perturbation}
\eeq
ith $\rho_{q}$ being a small amplitude. The variational free energy 
per monomer associated with this stretching can be expanded
in a power series in the amplitude $\rho_{q}$
\beq
	f= f_{0}+\rho_{q}^{2}f_{q}+\ldots
\eeq
Obviously, the uniformly stretched conformation does not
have the minimum energy if the $f_{q}$ is negative. 
The true minimum can be found by minimization over all possible stretchings,
but we will not attempt to solve this problem here. The 
results we obtain, however, show that when the instability occurs, 
it does so in a particular range of wavelengths, indicating that 
the minimum energy is obtained in a regular  necklace conformation, rather
than in a randomly spaced  one. 

	The linear stability analysis against small perturbations can also 
be carried out for perturbations that are transversal to the chain. 
These perturbations do not lead to instabilities but their study 
is important in understanding the persistence length of polyelectrolyte chains
\cite{Witten,Ha,Joanny} and will be studied elsewhere \cite{future}.

	The contribution to the free energy correction from the
entropic Gaussian part of the model Hamiltonian is simply
$(3/4)q^{2}\rho_{q}^{2}$. For the potential interaction part we 
write the contribution of a pair of monomers with labels $t'$ and $t$
($t' >t$) as
\beq
	\int \frac{d{\bf k} }{(2\pi)^{3/2}}\exp[-\frac{6(t'-t)k^{2}}{b'^{2}}] 
	\exp[i{\bf k}\cdot(\rho(t')-\rho(t))\hut]V({\bf k}). \label{qpot}
\eeq
We expand  the exponential factor into 
\begin{eqnarray}
\exp&&[i{\bf k}\cdot(\rho(t')-\rho(t))\hut]
	=\exp[i{\bf k}\cdot \rho_{0}(t'-t) \hut]  \nonumber \\
&(&1+i{\bf k\cdot \hut}\rho_{q}(\cos(qt')-\cos(qt))  \nonumber \\
&&-\frac{1}{2}\rho_{q}^{2}{\bf k\cdot \hut}^{2}	
(\cos(qt')-\cos(qt))^{2} +\ldots )
\end{eqnarray}
This expansion is substituted  into (\ref{qpot}), a new variable $s=t'-t$
is introduced and the integration over $t$
eliminates all oscillatory terms not depending on $s$ only, to obtain the final 
result, 
\begin{eqnarray}
	V_{q}\rho_{q}^{2} &=&
	\frac{1}{2}\rho_{q}^{2}
	\int_{0}^{\infty}ds 
	\int \frac{ d {\bf k}}{(2 \pi)^{3}} \nonumber \\
	&& ({\bf k} \cdot \hut)^{2}\left[ -1 + \cos( qs) \right]
	\exp[i{\bf k}\cdot \rho_{0}\hut]\hat{V} ({\bf k}).
\end{eqnarray}

	The criterion for instability is then:
\beq
	f_{q}=\frac{3}{4}q^{2}+V_{q} < 0. \label{crit}
\eeq
	Concrete evaluation of eq. (\ref{crit}) can be carried out
for specific values of $\bje$, $\kappa$ and $\nu$. The criteria is 
met for a bounded region of $q$ values. 
Among the $q$ values for which the system is unstable there will
be one value $q^{\*}$ for which the determinant function $f_{q}$ is 
most negative. 
We identify this wave-number with the inverse of the most likely
spacing between the beads of the necklace. 

	It is easier to discuss the results keeping $\kappa$ constant.
	For  a given value of $\kappa$ we can compute the region 
of stability in the $\bje$-$\nu$ plane and we present one such example 
in figure 3. On changign $\kappa$, the general morphology of the 
diagram is still the same as in figure 3 and our results
are consistent with those
 obtained by Dobrynin \etal \cite{Rub}. There are three main
regions. The collapsed state that occurs for sufficiently
large values of the attractive interaction disscused before, the rod-like 
conformation
region in which the electrostatic interaction is dominant, and the
more interesting part for our purposes, a narrow strip between them
which is still rodlike but has undergone a finite wave-length modulation.
In this last region, we have identified the points for which 
the value of the critical wavenumber $q^{\*}$ is the same. We also note
that the instability region for necklaces starts always at a finite value of
the Bjerrum number, since it is necessary for the system to have enough
electrostatic energy to elongate the chain.

\begin{figure}
\epsfxsize=2.5in
\epsfbox{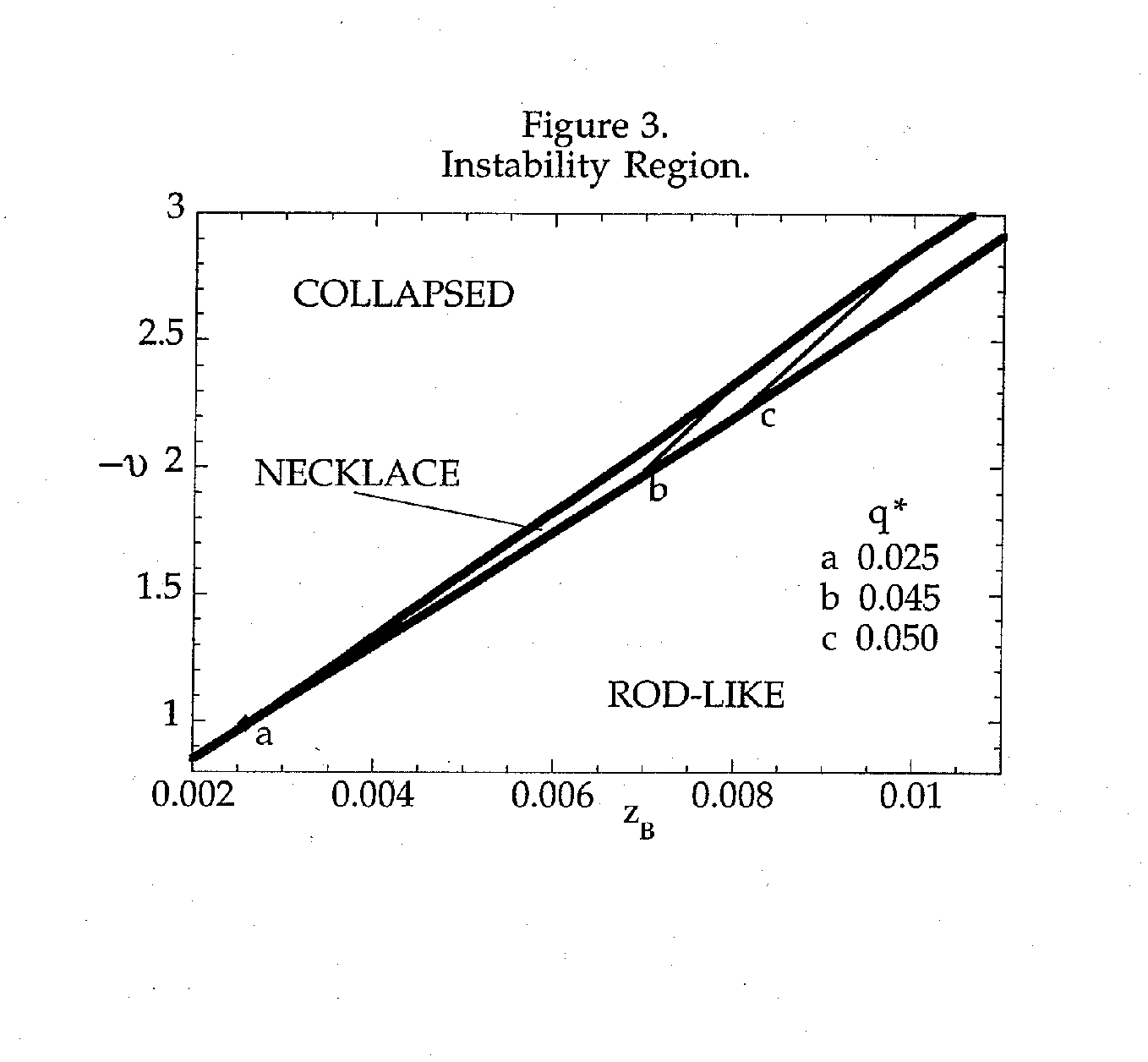}
\caption{ Region of neckalce instability for $\kappa=10^{-4}$ in the diagram $-\nu-\bje$.
Point {\it a} marks the limit of the 
necklace region, and lines {\it b} and {\it c} mark points of
equal values of the most unstable wave-numbers $q^{*}$. 
}
\end{figure}

	Different values of $\kappa$ produce the same general results in that 
we always reach, first,  a point of instability against collapse,
and for sufficiently large charge,  
and higher 
values of the attractive interaction, a necklace pattern. 
The main effect of a decreasing $\kappa$ is to make the electrostatic
interaction stronger, thus pushing the required values of $\nu$ for
each instability ever higher. 

 We add further detail to our example to illustrates another feature of the 
system. We have
computed for $\kappa=10^{-4}$ and $\bje=.001$ the region of instability 
and the region of unstable modes.
The results appear in figure 4. There,  we confirm the different nature
of the collapsed and the necklace state. The region on $q$ space 
which is unstable is always finite. To obtain a smooth transition to
the collapsed state we would need this region to grow and touch
large wavelength modes ($q \approx 0$) but this is not the case.

\begin{figure}
\epsfxsize=2.5in
\epsfbox{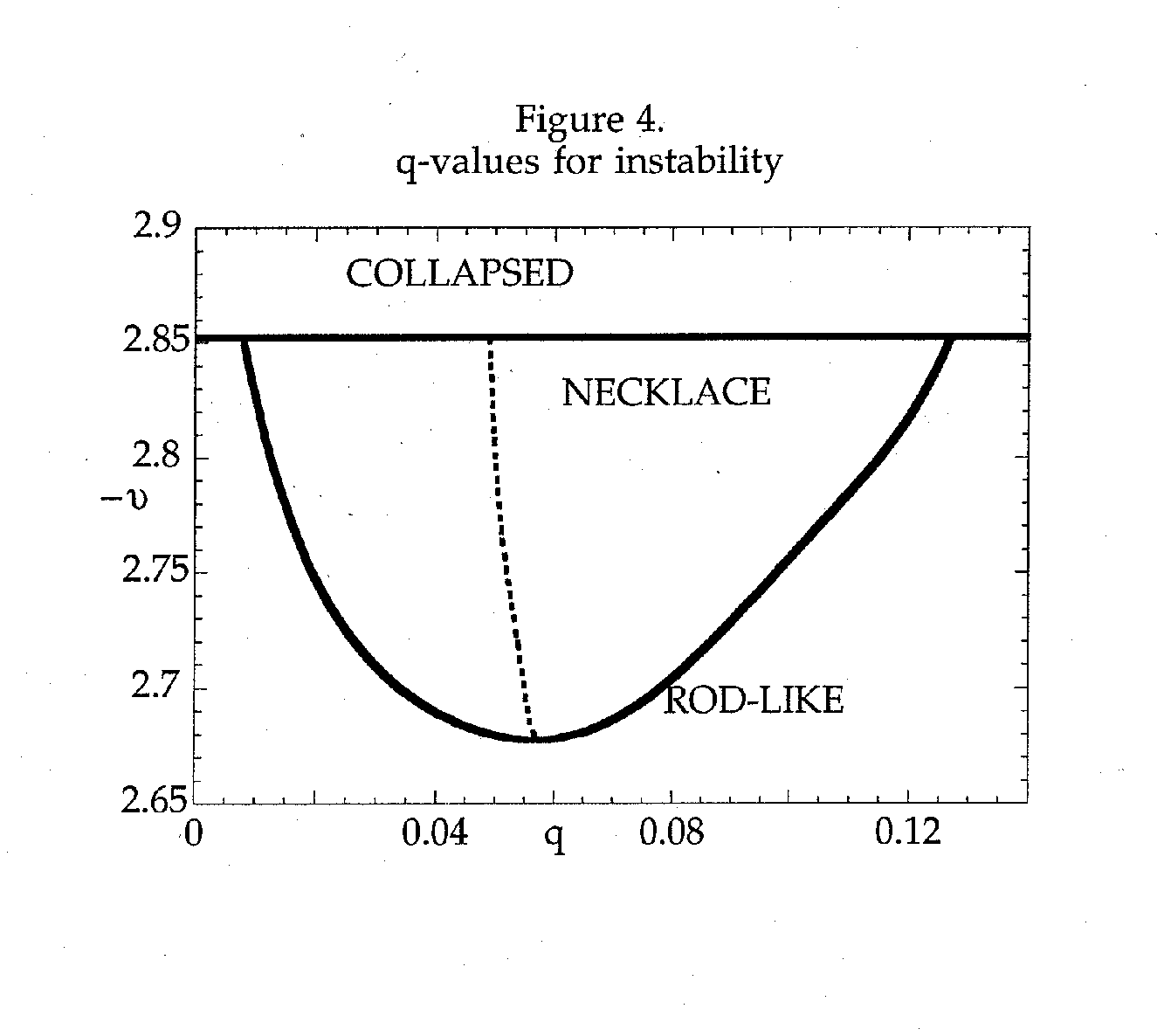}
\caption{Necklace unstable modes $q$ for $\kappa=10^{-4}$ and $\bje=0.01$. 
The dashed line marks the most unstable mode for each value of $\nu$. 
}
\end{figure}

\section{Conclusions}

	We confirm necklace formation in polyelectrolytes using the variational 
principle  technique. 
	In this calculational scheme we are limited to consider 
the regime in which the entropic free energy of the chain is still comparable 
with the energy scale imposed by the  charge interaction,  which leads to 
weak stretching. 
This calculation gives us  not only insight 
on the onset region of the phenomena, but it also provides a concrete 
parameter space window for the experimental search of the transition.
The present  calculation can be refined in many ways. As we have mentioned 
before, it is important to consider the effects of a fixed bond length,
and secondly, we can complicate the Gaussian correlations in the 
trial Hamiltonian so as to consider non-neighboring monomers correlations.

	In the future it will be very important to determine the 
exact location of the transition between the collapsed and extended
conformations.
Since the region where the necklace formation is observed  lies
between these two states, 
it might happen that  the 
collapsed state acquires the lowest energy before the onset of the finite
wavelenght instability. In that case the necklace cannot be observed. On the other hand,
this scenario should not be true in general since our method can also
underestimate the point at which the necklace transition occurs. Again, the
physical onset of the necklace preceeds the finite wavelenght 
instability of the rod-like conformation.

	Finally , from the methodological point of view, we believe that the
uses of the variational approach have not been exhausted. It
provides means to test the different structures arising in single
and multiple chains systems. For example, ona can study the case in which the 
charges are not quenched along the chain, a problem considered by Higgs and Joanny
\cite{Higgs} and recently extended to study interchain interactions by Liu and Ha \cite{Andrea}.

\section*{Acknowledgements}

	We thank the National Science Foundation for financial support 
under grant number DMR9509838.

\end{document}